\documentclass[conference]{IEEEtran}
\IEEEoverridecommandlockouts

\usepackage{amsmath,amssymb,amsfonts}
\usepackage{array}
\usepackage[caption=false,font=normalsize,labelfont=sf,textfont=sf]{subfig}
\usepackage{textcomp}
\usepackage{stfloats}
\usepackage{verbatim}
\usepackage{graphicx}
\usepackage{cite}
\usepackage{tabularx}
\usepackage{booktabs}

\def\BibTeX{{\rm B\kern-.05em{\sc i\kern-.025em b}\kern-.08em
    T\kern-.1667em\lower.7ex\hbox{E}\kern-.125emX}}

\usepackage{acronym}
\acrodef{5G}{the fifth generation}
\acrodef{MIMO}{multiple-input multiple-output}
\acrodef{RF}{radio frequency}
\acrodef{LoS}{line-of-sight}
\acrodef{NLoS}{non-line-of-sight}
\acrodef{AoA}{angle-of-arrival}
\acrodef{AoD}{angle-of-departure}
\acrodef{UPA}{uniform planar array}
\acrodef{ARV}{array response vector}
\acrodef{EM}{electromagnetic}
\acrodef{MA}{movable antenna}
\acrodef{ERA}{electromagnetically reconfigurable antenna}
\newtheorem{definition}{\textbf{Definition}}
\newtheorem{remark}{\textbf{Remark}}

\usepackage{color}

\usepackage{tikz} 
\usepackage[utf8]{inputenc}
\usepackage{pgfplots}
\usepackage{tabu,longtable}
\usepackage{diagbox}
\usepackage{threeparttable}
\usepackage{makecell}
\usepackage{multirow} % enable table multirow
\usepackage[bottom=1.04in,top=0.75in,left=0.625in,right=0.625in]{geometry}
\usepackage{units} 		% use \unit command
\usepackage{bm} 		% for bold symbol
\usepackage{amssymb} 	% for AMS symbols
\newcommand{\TT}{\mathsf{T}}

\newcommand{\av}{{\bf a}}
\newcommand{\bv}{{\bf b}}

\newcommand{\fv}{{\bf f}}
\newcommand{\gv}{{\bf g}}

\newcommand{\kv}{{\bf k}}

\newcommand{\pv}{{\bf p}}
\newcommand{\qv}{{\bf q}}

\newcommand{\xv}{{\bf x}}
\newcommand{\yv}{{\bf y}}

\newcommand{\Am}{{\bf A}}
\newcommand{\Bm}{{\bf B}}

\newcommand{\Xm}{{\bf X}}

\usepackage{hyperref} 	% for URL
\usepackage{amsmath}

\usepackage{algorithm}
\usepackage{algpseudocode}% http://ctan.org/pkg/algorithmicx
\algtext*{EndWhile}% Remove "end while" text
\algtext*{EndIf}% Remove "end if" text
\algtext*{EndFor}

\makeatletter
\algnewcommand{\LineComment}[1]{\Statex \hskip\ALG@thistlm \(\triangleright\) #1}
\makeatother
\begin{document}
\bstctlcite{IEEEexample:BSTcontrol} % Control reference style

\title{Enhanced Beampattern Synthesis Using Electromagnetically Reconfigurable Antennas\\ \vspace{-0.2em}
\author{
Pinjun Zheng, Md. Jahangir Hossain, Anas Chaaban\\
\textit{School of Engineering, University of British Columbia, Kelowna, Canada}\\
Email: \{pinjun.zheng; jahangir.hossain; anas.chaaban\}@ubc.ca \vspace{-0.5em}
} 
}

\maketitle

\begin{abstract}
Beampattern synthesis seeks to optimize array weights to shape radiation patterns, playing a critical role in various wireless applications. In addition to theoretical advancements, recent hardware innovations have facilitated new avenues to enhance beampattern synthesis performance. This paper studies the beampattern synthesis problem using newly proposed electromagnetically reconfigurable antennas (ERAs). By utilizing spherical harmonics decomposition, we simultaneously optimize each antenna's radiation pattern and phase shift to match a desired beampattern of the entire array. The problem is formulated for both far-field and near-field scenarios, with the optimization solved using Riemannian manifold techniques. The simulation results validate the effectiveness of the proposed solution and illustrate that ERAs exhibit superior beampattern synthesis capabilities compared to conventional fixed radiation pattern antennas. This advantage becomes increasingly significant as the array size grows.
\end{abstract}
\vspace{-0.7em}
\section{Introduction}

Since their inception in the 1990s, \ac{MIMO} technologies have served as a cornerstone of wireless communication systems~\cite{PAULRAJ2004Overview}. With the evolution of \ac{5G}, various reconfigurability features in \ac{MIMO} systems have garnered significant attention. This has driven the development of innovative radiating and reflecting surfaces and antennas, such as reconfigurable intelligent surfaces~\cite{Wang2024Wideband} and fluid antennas~\cite{Wong2021Fluid}. Recently, \acp{ERA} have emerged as another evolutionary extension of \ac{MIMO} technologies. This type of design enables the reconfigurability of each antenna element’s radiative properties—such as operating frequency, radiation pattern, and polarization direction—by controlling the connection states or geometric configurations of solid or liquid materials. A recent \ac{ERA} hardware is reported in~\cite{Wang2025Electromagnetically}. 

Broadly, the \ac{ERA} can be classified as a subset of fluid antennas, as fluid antennas refers to ``\emph{any software-controllable fluidic, conductive, dielectric structure, or even reconfigurable RF-pixels that can change its shape and position to reconfigure the gain, radiation pattern, operating frequency, and other characteristics}~\cite{Wong2022Bruce}''. However, most existing studies on fluid antennas have primarily focused on the spatially reconfigurable type, often referred to interchangeably as \acp{MA}~\cite{Zhu2024Historical}. These \acp{MA} can adapt to positions and/or orientations with favorable channel conditions, leveraging their flexibility to achieve enhanced spatial diversity gains. Nevertheless, they usually rely on intricate mechanical movement systems and high-precision motion control and optimization algorithms, whereas directly regulating the antenna's radiative properties offers a more efficient alternative. In this paper, \ac{ERA} refers specifically to antennas with reconfigurable radiation patterns, a class of systems also called reconfigurable massive \ac{MIMO}~\cite{Ying2024Reconfigurable} or electronically steerable passive array radiators (ESPAR)~\cite{Zhang2023Analog}.

In this work, we demonstrate the significant potential of the \ac{ERA} in addressing the \emph{beampattern synthesis problem}. In many wireless scenarios, it is critical to direct transmitter power toward specific target regions while minimizing signal leakage to eavesdroppers and mitigating interference among multiple access points or radar systems operating in the same area. In fact, beampattern synthesis has been extensively studied over the years~\cite{Stoica2008Waveform,He2011Wideband,Zhong2022RMOCG}. Numerous effective signal processing solutions, such as cyclic optimization~\cite{Stoica2008Waveform} and Riemannian manifold optimization~\cite{Zhong2022RMOCG}, have been proposed to address this problem. However, these methods, based on conventional antenna arrays, typically employ a fixed radiation pattern for each element (often isotropic), achieving the desired beampattern by adjusting tuners (e.g., baseband signals~\cite{He2011Wideband} and beamforming weights~\cite{Zhong2022RMOCG}) in the feeding network, without considering the reconfigurability of antennas. Enabled by the ERA, beampattern synthesis performance can be further enhanced by leveraging the additional degrees of freedom introduced by the \ac{EM} reconfigurability of the antenna.

This paper formulates and solves the beampattern synthesis problem using the \ac{ERA}, which simultaneously configures element radiation patterns and phase shift reconfigurability. We model the radiation patterns of the ERA using spherical harmonics decomposition, enabling tractable control over each element's radiation pattern by adjusting the harmonic coefficients. Based on this model, we formulate an optimization problem for array beampattern synthesis and propose a Riemannian manifold-based optimization framework to solve it, yielding a pattern that closely approximates the desired beampattern. The proposed modeling, analysis, and optimization framework apply to both far-field and near-field scenarios.

\section{System Model}\label{sec:SM}

\begin{figure*}[t]
  \centering
  \includegraphics[width=0.9\linewidth]{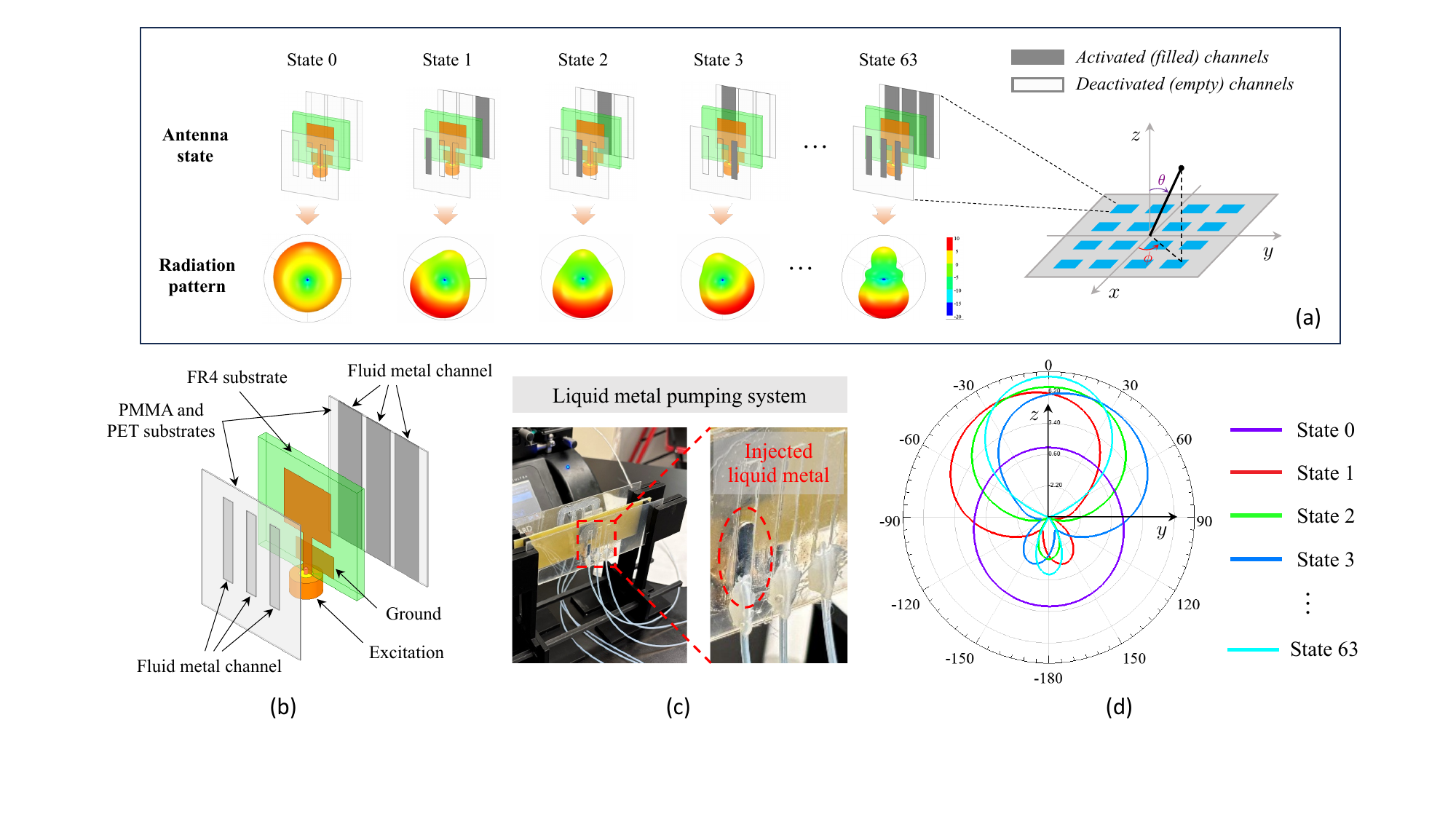}
  \vspace{-0.8em}
  \caption{
    An example of an \ac{ERA} array proposed in~\cite{Wang2025Electromagnetically}. (a) Array geometry and the radiation pattern reconfiguration mechanism. (b) Detailed antenna element structure, where each fluid metal channel can be filled or emptied with liquid metal thus enabling the radiation pattern reconfigurability. (c) A fabricated \ac{ERA} element prototype. (d) Sample radiation pattern plots in the $YOZ$ plane.
  }
  \label{fig_geo}
  \vspace{-0.7em}
\end{figure*} 

We consider a $N_\mathsf{x}\times N_\mathsf{y}$ transmitter \ac{UPA} deployed on the $XOY$ plane, where $N_\mathsf{x}$ and $N_\mathsf{y}$ denote the array dimensions along the $X$-axis and $Y$-axis, respectively. Let $N=N_\mathsf{x}N_\mathsf{y}$. We assume $N$ analog phase shifters are connected to this antenna array, and each antenna unit can further adjust its radiation pattern, as shown in Fig.~\ref{fig_geo}. This radiation pattern reconfigurability can be achieved through various techniques. Fig.~\ref{fig_geo} illustrates an example where liquid metal is used to reshape the radiative structure of each antenna, the detailed design of which can be found in~\cite{Wang2025Electromagnetically}.

\subsection{Spherical Harmonics Representation of Radiation Pattern}
The radiation pattern is essentially a signal on 2-sphere $\mathbb{S}^2$, which is defined as the locus of points in $\mathbb{R}^3$ with unit form, i.e.,  $\mathbb{S}^2\triangleq \{\qv\in\mathbb{R}^3|\qv^\TT\qv=1\}$.
In Euclidean space $\qv=[q_x,q_y,q_z]^\TT$, points on $\mathbb{S}^2$ are also conventionally parameterized by the inclination angle~$\theta$ and azimuth angle~$\phi$ of a spatial direction, as depicted in Fig.~\ref{fig_geo}. These two representations are equivalent and are related through the conversion relations:
$
    q_x = \sin{\theta}\cos{\phi},\ q_y = \sin{\theta}\sin{\phi},\ q_z=\cos{\theta}.
$

Using the angular representation, an antenna's radiation pattern can be decomposed into a superposition of infinitely many spherical harmonics~\cite{Xu20173D}. Assuming linear polarization, we can express an arbitrary amplitude field as~\cite{Costa2010Unified,Dokmanic2016Sampling,Ying2024Reconfigurable}
\begin{equation}\label{eq:SH}
    g(\theta,\phi) = \sum_{\ell=0}^{+\infty}\sum_{m=-\ell}^\ell b_{\ell m} Y_\ell^m(\theta,\phi).
\end{equation}
Here, $b_{\ell m}$ denotes the harmonic coefficient and $Y_\ell^m(\theta,\phi)$ is the \emph{real} spherical harmonics defined as
\begin{equation}\label{eq:Ylm}
Y_{\ell}^m(\theta,\phi)\!=\! \left\{
\begin{array}{ll}
\!\!\!\sqrt{2}N_\ell^mP_\ell^m(\cos{\theta})\cos{(m\phi)}, & m>0,\\
\!\!\!\sqrt{2}N_\ell^{|m|}P_\ell^{|m|}(\cos{\theta})\sin{(|m|\phi)}, & m< 0,\\
\!\!\!N_\ell^0P_\ell^0(\cos{\theta}), & m=0.
\end{array} \right.
\end{equation}
where $N_\ell^m=\sqrt{\frac{2\ell+1}{4\pi}\frac{(\ell-m)!}{(\ell+m)!}}$ is a normalization factor, and $P_\ell^m(\cos{\theta})$ represents the associated Legendre functions of $\ell^\text{th}$ degree (or level) and $m^\text{th}$ order (or mode). These infinitely many functions, $Y_{\ell}^m(\theta,\phi)$, $\ell\in[0,+\infty],\ m\in [-\ell,+\ell]$, constitute a complete real orthonormal basis on $\mathbb{S}^2$. 

While~\eqref{eq:SH} holds rigorously with an infinite number of coefficients $b_{\ell m}$, a practical truncation can be applied for operational convenience. An analysis of truncation errors for various antenna radiation patterns under different numbers of retained coefficients is provided in~\cite{Ying2024Reconfigurable}. In general, these truncation errors become negligible when a sufficiently large number of harmonic coefficients is used. 

\subsection{Far-Field Beampattern}
In the far-field, the \ac{EM} signal propagation is described as a planar wave, where the array response varies over the \acp{AoD}~$\{\theta,\phi\}$. 
We model the analog phase shifting feeding network as $\fv\in\mathbb{C}^{N}$, whose entries are given by $f_n=e^{j\beta_n}$, with $\beta_n$ denoting the phase shift corresponding to the~$n^\text{th}$~antenna element. Hence, the far-field beampattern of the \ac{ERA} array is calculated as~\cite{Balanis2015Antenna,Zheng2024Mutual}
\begin{equation}\label{eq:E}
    E(\theta,\phi) = \big|\big(\gv(\theta,\phi)\odot\av(\theta,\phi)\big)^\TT\fv\big|,
\end{equation}
where $\gv(\theta,\phi)=[g_1(\theta,\phi),g_2(\theta,\phi),\dots,g_N(\theta,\phi)]^\TT\in\mathbb{R}^{N}$ models the antenna gains, and $\av(\theta,\phi)\in\mathbb{C}^{N}$ is the \ac{ARV}, corresponding to the \ac{AoD} $\{\theta,\phi\}$. Here, $g_n(\theta,\phi)$ denotes the radiation pattern of the $n^\text{th}$ antenna, and the \ac{ARV} $\av(\theta,\phi)$ is defined as
\begin{equation}\label{eq:ARVFF}
    \av(\theta,\phi) \triangleq e^{-j2\pi\varphi_\mathsf{x}\kv(N_\mathsf{x})} \otimes e^{-j2\pi\varphi_\mathsf{y}\kv(N_\mathsf{y})}.
\end{equation}
where $\kv(n) = [0,1,\dots,n-1]^\TT$, $\varphi_\mathsf{x}=\frac{d}{\lambda}\sin{\theta}\sin{\phi}$ and $\varphi_\mathsf{y}=\frac{d}{\lambda}\sin{\theta}\cos{\phi}$  denote the spatial angles corresponding to the $X$ and $Y$ dimension, respectively, $d$ denotes the inter-element spacing of the array, and $\lambda$ denotes the signal wavelength. 
\begin{remark}
    The antenna gain is defined as the radiation intensity of an antenna in a given direction relative to that of an isotropic radiator~\cite{6758443}. The radiation density of an isotropic radiator can be computed as $\frac{P}{4\pi}$, where $P$ is the total radiated power and $4\pi$ represents the total solid angle of a sphere. Since this work focuses on beampattern synthesis, we omit this constant factor $\frac{P}{4\pi}$ as it is irrelevant to the array beampattern, assuming all antennas have the same radiated power.
\end{remark}

By defining the truncation length $T=L^2+2L+1$, we can approximate the radiation pattern of the $n^\text{th}$ antenna as
\begin{align}\label{eq:gn}
    g_n(\theta,\phi) \approx \sum_{\ell=0}^{L}\sum_{m=-\ell}^\ell b_{\ell m}^{(n)} Y_\ell^m(\theta,\phi) = \sum_{t=1}^T \tilde{b}_t^{(n)} \tilde{Y}_t(\theta,\phi),
\end{align}
where $\tilde{b}_t^{(n)} = b_{\ell m}^{(n)}$ and $\tilde{Y}_t(\theta,\phi) = Y_\ell^m(\theta,\phi)$, for $t=\ell^2+\ell+m+1$, $\ell\in[0,L]$, $m\in[-\ell,\ell]$. For convenience, we further concatenate 
\begin{align}
\bv_n&\triangleq[\tilde{b}_1^{(n)},\tilde{b}_2^{(n)},\dots,\tilde{b}_T^{(n)}]^\TT\in\mathbb{R}^T,\label{eq:bnFF}\\
\yv_n(\theta,\phi)&\triangleq[\tilde{Y}_1(\theta,\phi),\tilde{Y}_2(\theta,\phi),\dots,\tilde{Y}_T(\theta,\phi)]^\TT\in\mathbb{R}^T. \label{eq:ynFF}
\end{align}
Although here $\yv_n(\theta,\phi)$ is independent of the antenna index $n$, we retain this subscript for consistency in the later near-field extension. Based on~\eqref{eq:bnFF} and~\eqref{eq:ynFF}, we can rewrite~\eqref{eq:gn} as
\begin{equation}\label{eq:gn2}
    g_n(\theta,\phi)\approx \yv_n^\TT(\theta,\phi) \bv_n.
\end{equation}
Substituting~\eqref{eq:gn2} to~\eqref{eq:E} gives
\begin{equation}\label{eq:Ecompact}
    E(\theta,\phi) = |\bv^\TT\Am(\theta,\phi)\fv|,
\end{equation}
where $\bv=[\bv_1^\TT,\bv_2^\TT,\dots,\bv_N^\TT]^\TT\in\mathbb{R}^{NT}$, and  $\Am(\theta,\phi)\in\mathbb{C}^{NT\times N}$ is a \ac{EM}-domain array response matrix~\cite{Ying2024Reconfigurable} expressed as
\begin{equation}\label{eq:A}
    \Am(\theta,\phi) = \mathbf{1}\av^\TT(\theta,\phi)\odot\text{blkdiag}\{\yv_1(\theta,\phi),\dots,\yv_N(\theta,\phi)\}.
\end{equation}
Here,~$\mathbf{1}$ represents an $NT$-length all-ones column vector, and blkdiag$\{\cdot\}$ creates a block diagonal matrix from its inputs with zeros in off-diagonal blocks.
From~\eqref{eq:Ecompact}, we observe that the array beampattern is influenced by (i) the spherical harmonics coefficient vector $\bv$, which governs the radiation pattern of each \ac{ERA} element, and (ii) the phase shift vector $\fv$ of the analog feeding network.

\subsection{Near-Field Beampattern}
In this work, we refer to the near-field region as the region between the \emph{direction-independent uniform-power distance} and the \emph{Rayleigh distance}, where the uniform spherical wave~(USW) model can be adopted to describe signal propagation~\cite{Lu2024Tutorial}. In this region, the \ac{ARV} depends not only on the \ac{AoD} but also on the propagation distance, i.e., the exact position of the receiver. According to the USW model, the \ac{ARV} in~\eqref{eq:ARVFF} can be rewritten as~\cite{Lu2024Tutorial}
\begin{equation}\label{eq:ARVNF}
    \av(\pv_\mathsf{R})\triangleq\big[e^{-j\frac{2\pi}{\lambda}\|\pv_1-\pv_\mathsf{R}\|},\dots,e^{-j\frac{2\pi}{\lambda}\|\pv_N-\pv_\mathsf{R}\|}\big]^\TT\!\!\in\mathbb{C}^N,
\end{equation}
where $\pv_\mathsf{R}$ denotes the position of the receiver, and $\pv_n$ denotes the position of the $n^\text{th}$ antenna element, for $n=1,2,\dots,N$.

In the near-field, the spherical harmonics basis vector~$\yv_n$ varies across different antenna elements due to the changing \ac{AoD} from each element to the target near-field receiver. Therefore, we can rewrite~\eqref{eq:ynFF} as
\begin{equation}\label{eq:ynNF}
\yv_n(\pv_\mathsf{R})\triangleq[\tilde{Y}_1(\theta_n,\phi_n),\tilde{Y}_2(\theta_n,\phi_n),\dots,\tilde{Y}_T(\theta_n,\phi_n)]^\TT,
\end{equation}
where the \ac{AoD} $\theta_n = \arccos{([({\pv_\mathsf{R}-\pv_n})/{\|\pv_\mathsf{R}-\pv_n\|}]_3)},\ 
    \phi_n = \arctan2{([{\pv_\mathsf{R}-\pv_n}]_2,[{\pv_\mathsf{R}-\pv_n}]_1)}.$
Then, the near-field beampattern can be calculated by (i) substituting~\eqref{eq:ARVNF} and~\eqref{eq:ynNF} into~\eqref{eq:A} to obtain the near-field \ac{EM}-domain array response $\Am(\pv_\mathsf{R})$ and (ii) applying~\eqref{eq:Ecompact} to compute the near-field beampattern as $E(\pv_\mathsf{R}) = |\bv^\TT \Am(\pv_\mathsf{R}) \fv|$.

\section{Beampattern Synthesis}
Based on the model derived in Section~\ref{sec:SM}, we now formulate and solve the beampattern synthesis problem.

\subsection{Beampattern Synthesis Problem Formulation}
Suppose a desired beampattern represented by $S$ real-value samples $\{D_s\}_{s=1}^S$. In far-field scenarios, these samples are defined in the angle domain, i.e., $\{D(\theta_s,\phi_s)\}_{s=1}^S$; while in the near-field, these samples are defined in the position domain, i.e., $\{D(\pv_{\mathsf{R},s})\}_{s=1}^S$. Each of these target sample corresponds to an \ac{EM}-domain array response matrix: $\Am_s = \Am(\theta_s, \phi_s)$ for the far-field case, or $\Am_s = \Am(\pv_{\mathsf{R},s})$ for the near-field case. We can then formulate the beampattern synthesis problem as
\vspace{-0.8em}
\begin{equation}\label{eq:opt1}
\begin{aligned}
    \min_{\bv\in\mathbb{R}^{NT},\fv\in\mathbb{C}^N}&\ \sum_{s=1}^S\ \omega_s(D_s-|\bv^\TT\Am_s\fv|)^2,\\
    \text{s.t.\ \ \quad}&\ \|\bv_n\|^2 = P,\ |f_n|=1,\ n=1,2,\dots,N, 
\end{aligned}
\end{equation}
where $\omega_s$ is a given weight, and $P$ denotes the total radiated power of each antenna.

To overcome the intractability introduced by the modulus operation in the objective function of~\eqref{eq:opt1}, we introduce $S$ auxiliary variables $\psi_s \in \mathbb{R},\ s = 1, 2, \dots, S$~\cite{He2011Wideband}. As a result, the optimization problem becomes equivalent to
\begin{equation}\label{eq:opt2}
\begin{aligned}
    \min_{\scriptsize \begin{array}{c} \bv\!\in\!\mathbb{R}^{NT}\!\!,\fv\!\in\!\mathbb{C}^N\!\!,\\ \{\psi_s\}_{s=1}^S\in\mathbb{R}^S\end{array}}&\ \sum_{s=1}^S\ \omega_s\big|D_s e^{j\psi_s}-\bv^\TT\Am_s\fv\big|^2,\\
    \text{s.t.\ \qquad}&\ \|\bv_n\|^2 = P,\ |f_n|=1,\ n=1,2,\dots,N. 
\end{aligned}
\end{equation}
Given that $\bv$ and $\fv$ are constrained variables, while $\{\psi_s\}_{s=1}^S$ are unconstrained, we employ the \emph{block-coordinate descent method}, also referred to as the \emph{non-linear Gauss-Seidel method} or \emph{alternating minimization}~\cite{Grossi2023Beampattern}, for operational convenience. Let $k$ index the iteration. At each iteration~$k$, given previous candidate $\bv^{(k-1)}$, $\fv^{(k-1)}$, and $\{\psi_s^{(k-1)}\}_{s=1}^S$, we first update the auxiliary valuables as
\begin{equation}\label{eq:psis}
    \psi_s^{(k)} = \arg\big(\big(\bv^{(k-1)}\big)^\TT\Am_s\fv^{(k-1)}\big) 
\end{equation}
Then, we fix $\psi_s^{(k)}$ and update $\bv^{(k)}$ and $\fv^{(k)}$ by solving the following constrained optimization problem:
\begin{equation}\label{eq:opt3}
\begin{aligned}
    \min_{\scriptsize  \bv\in\mathbb{R}^{NT}\!\!,\fv\in\mathbb{C}^N}&\ \sum_{s=1}^S\ \omega_s\big|D_s e^{j\psi_s^{(k)}}-\bv^\TT\Am_s\fv\big|^2,\\
    \text{s.t.\ \quad}&\ \|\bv_n\|^2 = P,\ |f_n|=1,\ n=1,2,\dots,N. 
\end{aligned}
\end{equation}

\subsection{Solving~\eqref{eq:opt3} Using Riemannian Manifold Optimization}
The main challenges in~\eqref{eq:opt3} arise from the constraints. To preserve these constraints throughout the optimization, we leverage Riemannian manifold techniques. First, we introduce the definitions of the oblique manifold $\mathcal{OB}$ and the complex circle manifold $\mathcal{S}$~\cite{Boumal2023Introduction} as follows.
\begin{definition}
    The oblique manifold $\mathcal{OB}(m,n)$ is defined as the set of all $m\times n$ real matrices whose columns have unit Euclidean norm, i.e., $\mathcal{OB}(m,n)=\{\mathbf{X}\in\mathbb{R}^{m\times n}:\text{ddiag}(\Xm^\TT\Xm)=\mathbf{I}_n\}$. Here, $\text{ddiag}(\cdot)$ returns a diagonal matrix whose diagonal elements are those of the matrix in the argument.
\end{definition}
\begin{definition}
    The complex circle manifold $\mathcal{S}(n)$ is defined as a $n$-length complex vector whose each entries have unit amplitude, i.e., $\mathcal{S}(n)=\{\xv\in\mathbb{C}^n:|x_\ell|=1,\ell=1,2,\dots,n\}$.
\end{definition}

Based on these two definitions, one can rewrite~\eqref{eq:opt3} into 
\begin{equation}\label{eq:opt4}
\begin{aligned}
    \min_{\scriptsize \begin{array}{c} \Bm\!\in\!\mathcal{OB}(T,\!N),\\ \fv\!\in\!\mathcal{S}(N)\end{array}}&\ \sum_{s=1}^S \!\omega_s\big|D_s e^{j\psi_s^{(k)}}\!\!-{\sqrt{P}}{\text{vec}(\Bm)}^\TT\!\Am_s\fv\big|^2,
\end{aligned}
\end{equation}
which can be effectively solved using the Riemannian manifold optimization tools. Since the product of two embedded submanifolds is still a Riemannian manifold~\cite{Boumal2023Introduction}, one can solve~\eqref{eq:opt3} in a more compact way by defining a compound manifold as~\cite{Zheng2024LEO} $(\Bm,\fv)\in\mathcal{M}\triangleq\mathcal{OB}(T,N)\times\mathcal{S}(N)$, where $\times$ denotes the Cartesian product.

Similar to the gradient descent algorithm in Euclidean space, optimization over a Riemannian manifold is implemented by using the \emph{Riemannian gradient}~\cite{Boumal2023Introduction,Zheng2024Coverage}. Note that within each iteration~$k$, we introduce a new inner loop, indexed by $i$, to solve~\eqref{eq:opt4}. At iteration $i$, given the previous candidate variable~$(\Bm^{(i-1)},\fv^{(i-1)})$, we first compute the Riemannian gradient by projecting the Euclidean gradient onto the tangent space $\mathcal{T}_{(\Bm^{(i-1)},\fv^{(i-1)})}\mathcal{M}$ of the manifold $\mathcal{M}$. We then update the optimization variable in the direction of the Riemannian gradient and retract it from the tangent space $\mathcal{T}_{(\Bm^{(i-1)},\fv^{(i-1)})}\mathcal{M}$ onto the manifold~$\mathcal{M}$, thus obtaining new candidate variable~$(\Bm^{(i)},\fv^{(i)})$. This ensures that the iterations always proceed within the defined manifold. The detailed expressions for the projection from the Euclidean space to the tangent space and the retraction from the tangent space to the manifold, for both the oblique and complex circle manifolds, can be found in, e.g.,\cite{Selvan2012Descent,Alhujaili2019Transmit,Boumal2023Introduction}. These algorithms have been effectively implemented and integrated into toolboxes such as Manopt~\cite{manopt}.

\section{Simulation Results}

\begin{figure}[t]
  \centering
  \includegraphics[width=1\linewidth]{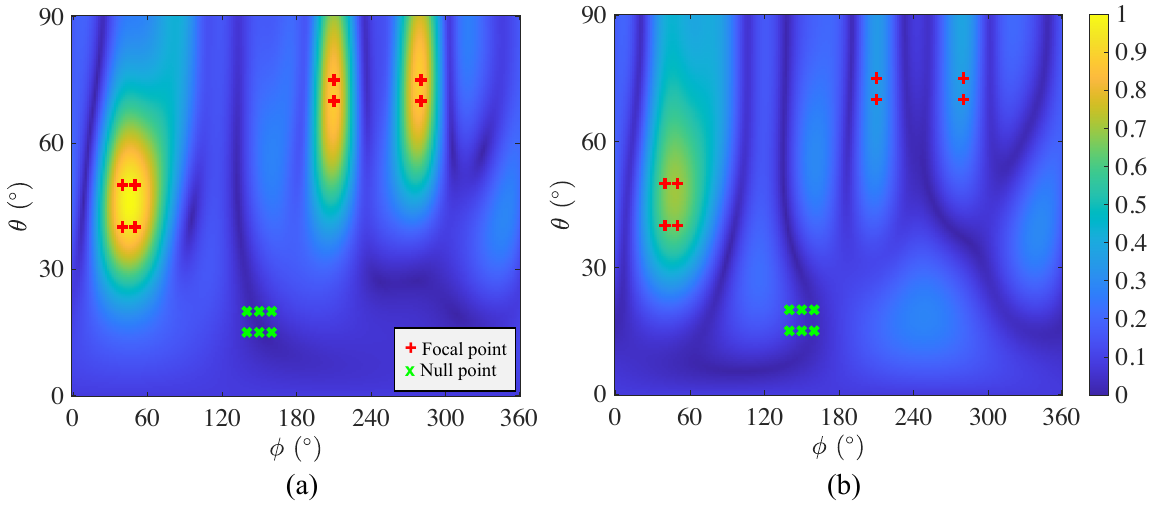}
  \vspace{-2em}
  \caption{
    Far-field beampattern synthesized using the array of (a)~\acp{ERA} and (b)~isotropic antennas.
  }
  \label{fig_FF}
  \vspace{-0.5em}
\end{figure} 

\subsubsection{Simulation Setup}
We simulate a signal with a frequency in the mmWave band at \unit[30]{GHz}. In far-field scenarios, the transmitter is configured with a $4 \times 4$ antenna array, while in near-field scenarios, it is set as a $64 \times 1$ array along the $Y$-axis. These antenna arrays can be composed of ERAs or fixed isotropic antennas. Both cases will be evaluated and compared to assess their performance. The antenna spacing is set to half the wavelength. For \ac{ERA}, the spherical harmonics truncation degree is set to $L=4$, and the antenna radiation patterns and phase shifters are optimized by applying~\eqref{eq:psis} and solving~\eqref{eq:opt4} alternately. For the isotropic antenna benchmark, only phase shifters are optimized by using the complex circle manifold optimization.

\begin{remark}
    The harmonics expression defined in~\eqref{eq:Ylm} can produce negative values, while the antenna radiation pattern must be positive. To maintain this constraint, a simple approach is to fix a sufficiently large $0^\text{th}$-degree coefficient~$b_{00}$ and optimize only the remaining coefficients $b_{\ell m},\ \ell \in [1,L],\ m \in [-\ell,\ell]$.
\end{remark}

\begin{remark}
   Although this paper optimizes the antenna’s radiation pattern arbitrarily, which may be impractical, the method remains valuable from a signal-processing perspective. For example, one potential application of this result is to project the optimized pattern onto a set of available radiation patterns that can be realized by the \ac{ERA}, thereby identifying the optimal available radiation pattern. However, we leave this for future investigation due to the space limitation.
\end{remark}

\begin{figure}[t]
  \centering
  \includegraphics[width=1\linewidth]{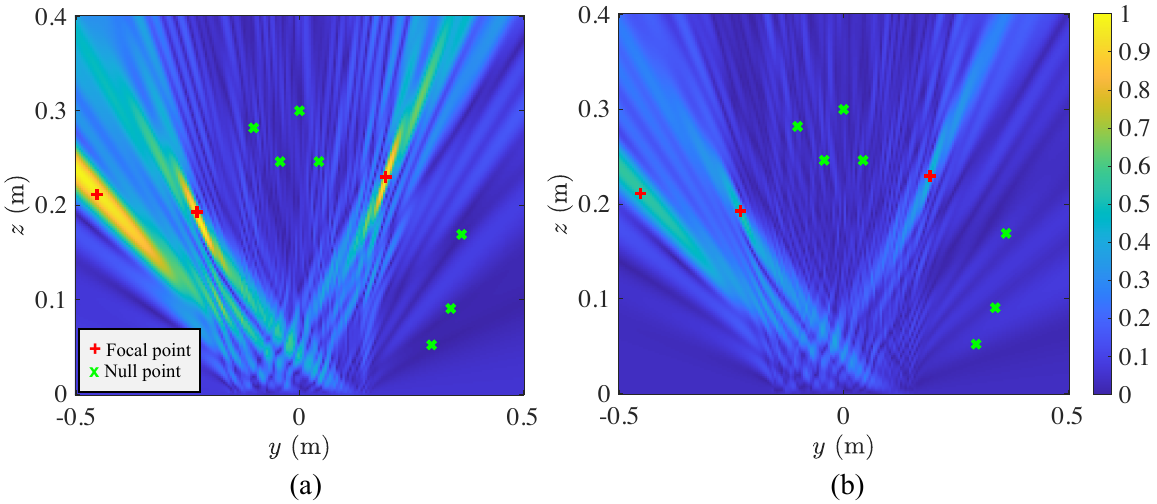}
  \vspace{-2em}
  \caption{
    Near-field beampattern synthesized using the array of (a)~\acp{ERA} and (b)~isotropic antennas.
  }
  \label{fig_NF}
  \vspace{-0.5em}
\end{figure} 

\begin{figure}[t]
  \centering
  \includegraphics[width=\linewidth]{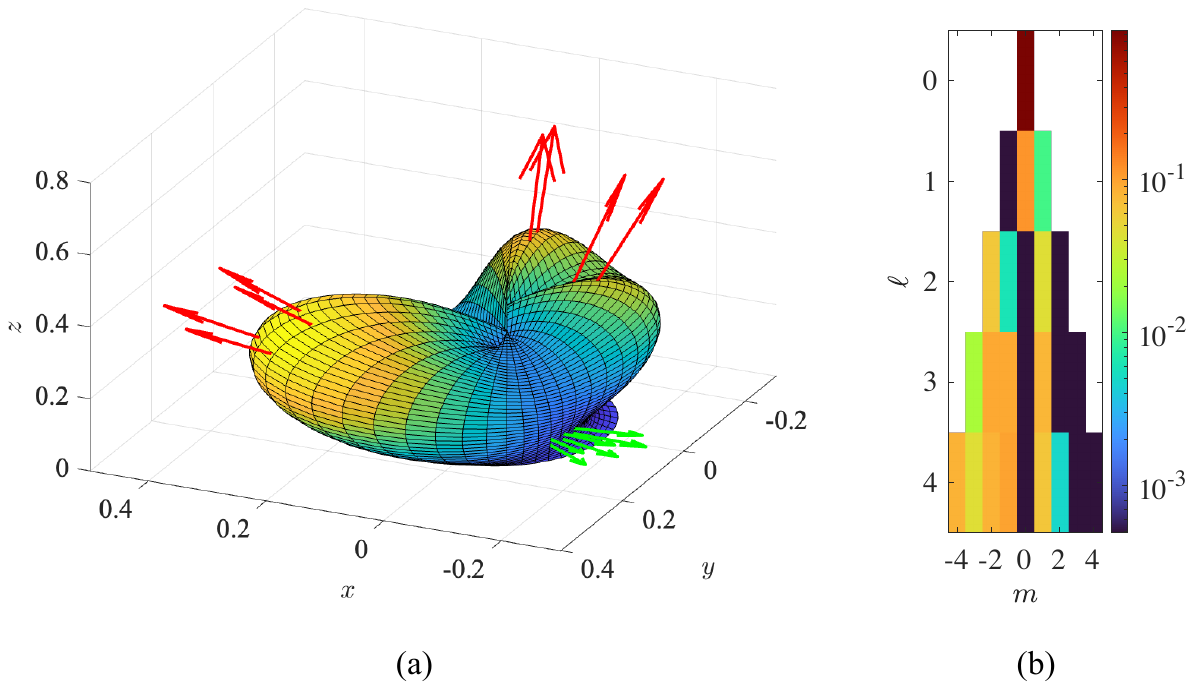}
  \vspace{-2em}
  \caption{
  Visualization of (a) the optimized radiation pattern of the first antenna in the array in Fig.~\ref{fig_FF}, where red and green arrows indicate the desired focal and nulling directions, respectively, and (b) the corresponding coefficients of the real spherical harmonics.}
  \label{fig_pattern}
\end{figure} 

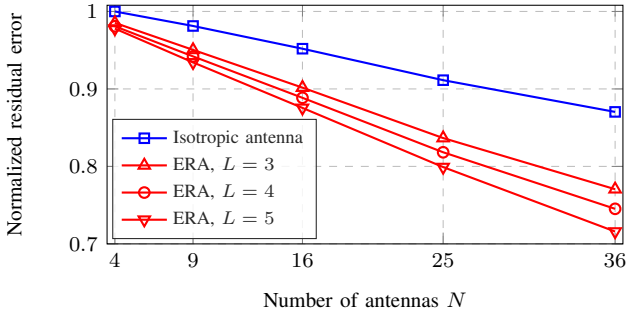
\begin{figure}[t]
    \centering
\begin{tikzpicture}

\begin{axis}[%
width=2.7in,
height=1.25in,
at={(0in,0in)},
scale only axis,
xmin=3.5,
xmax=36.5,
xtick={ 4,  9, 16, 25, 36},
xlabel style={font=\color{white!15!black},font=\footnotesize},
xticklabel style = {font=\color{white!15!black},font=\footnotesize},
xlabel={Number of antennas $N$},
ymin=0.7,
ymax=1.008,
ylabel style={font=\color{white!15!black},font=\footnotesize},
yticklabel style = {font=\color{white!15!black},font=\footnotesize},
ylabel={Normalized residual error},
axis background/.style={fill=white},
xmajorgrids,
ymajorgrids,
grid style={dashed},
legend style={at={(0.01,0.02)}, anchor=south west, font=\scriptsize, legend cell align=left, align=left, draw=white!15!black, fill opacity=0.85}
]
\addplot [color=blue, line width=0.8pt, mark=square, mark options={solid, blue}, mark size=1.8pt]
  table[row sep=crcr]{%
4	1\\
9	0.981259286716615\\
16	0.951914449450974\\
25	0.91129824528115\\
36	0.870347844053863\\
};
\addlegendentry{Isotropic antenna}

\addplot [color=red, line width=0.8pt, mark=triangle, mark options={solid, red}, mark size=2.5pt]
  table[row sep=crcr]{%
4	0.985237421361628\\
9	0.950420677069101\\
16	0.901691738491147\\
25	0.836595439226574\\
36	0.770569415023614\\
};
\addlegendentry{ERA, $L=3$}

\addplot [color=red, line width=0.8pt, mark=o, mark options={solid, red}]
  table[row sep=crcr]{%
4	0.980817380124508\\
9	0.941991444344061\\
16	0.888545134713093\\
25	0.818243576433974\\
36	0.745248730513361\\
};
\addlegendentry{ERA, $L=4$}

\addplot [color=red, line width=0.8pt, mark=triangle, mark options={solid, rotate=180, red}, mark size=2.5pt]
  table[row sep=crcr]{%
4	0.977884951240174\\
9	0.934270688006226\\
16	0.875522816531246\\
25	0.799118914797423\\
36	0.715929009962697\\
};
\addlegendentry{ERA, $L=5$}

\end{axis}
\end{tikzpicture}%
    \vspace{-3em}
    \caption{Evaluation of the far-field optimization residual errors versus the number of antennas for different spherical harmonics truncation degrees $L$.}
    \label{fig_resErr}
    \vspace{-1em}
\end{figure}

\subsubsection{Results Analysis}
Fig.~\ref{fig_FF} and Fig.~\ref{fig_NF} present the far-field and near-field beampattern synthesis results, respectively. In each scenario, we compare the synthesized beampatterns using the \ac{ERA} array and fixed isotropic antenna array. The desired beampattern is defined by a set of focal points (shown as red markers) and null points (shown as green markers). It is evident that in both far-field and near-field scenarios, the \ac{ERA} significantly outperforms the isotropic antenna, directing more power to the focal points while more effectively mitigating power at the null points. This also confirms the effectiveness of the proposed optimization approach. An intuitive demonstration of the optimized radiation pattern and corresponding harmonic coefficients of a selected antenna in the far-field trial is shown in Fig.~\ref{fig_pattern}. Additionally, Fig.~\ref{fig_resErr} evaluates the normalized residual errors of the optimization (i.e., the objective function value of~\eqref{eq:opt1}) versus the number of antennas for different spherical harmonics truncation degrees $L$. We observe that the performance improvement provided by the \ac{ERA} increases with the array size. Moreover, a higher truncation degree $L$ further enhances the beampattern synthesis capability, though at the cost of increased computational complexity.

\section{Conclusion}
This paper investigates beampattern synthesis using \acp{ERA}, which offers a significantly higher degree of freedom by simultaneously adjusting the radiation pattern and phase shift of each antenna element. By leveraging spherical harmonics representation, we derive the beampattern of an ERA array in both far-field and near-field scenarios and formulate the beampattern synthesis problem accordingly. This constrained optimization problem is effectively addressed using Riemannian manifold optimization techniques. Simulation results confirm the effectiveness of the proposed approach and demonstrate the substantial potential of ERAs in beampattern synthesis. These findings pave the way for future research on integrating ERA technology into advanced wireless communication systems and various other electromagnetic applications.

\bibliography{references}

\begin{thebibliography}{10}
\providecommand{\url}[1]{#1}
\csname url@samestyle\endcsname
\providecommand{\newblock}{\relax}
\providecommand{\bibinfo}[2]{#2}
\providecommand{\BIBentrySTDinterwordspacing}{\spaceskip=0pt\relax}
\providecommand{\BIBentryALTinterwordstretchfactor}{4}
\providecommand{\BIBentryALTinterwordspacing}{\spaceskip=\fontdimen2\font plus
\BIBentryALTinterwordstretchfactor\fontdimen3\font minus
  \fontdimen4\font\relax}
\providecommand{\BIBforeignlanguage}[2]{{%
\expandafter\ifx\csname l@#1\endcsname\relax
\typeout{** WARNING: IEEEtran.bst: No hyphenation pattern has been}%
\typeout{** loaded for the language `#1'. Using the pattern for}%
\typeout{** the default language instead.}%
\else
\language=\csname l@#1\endcsname
\fi
#2}}
\providecommand{\BIBdecl}{\relax}
\BIBdecl

\bibitem{PAULRAJ2004Overview}
A.~Paulraj, D.~Gore \emph{et~al.}, ``An overview of {MIMO} communications--a
  key to gigabit wireless,'' \emph{Proceedings of the IEEE}, vol.~92, no.~2,
  pp. 198--218, 2004.

\bibitem{Wang2024Wideband}
R.~Wang, Y.~Yang \emph{et~al.}, ``A wideband reconfigurable intelligent surface
  for {5G} millimeter-wave applications,'' \emph{IEEE Transactions on Antennas
  and Propagation}, vol.~72, no.~3, pp. 2399--2410, 2024.

\bibitem{Wong2021Fluid}
K.-K. Wong, A.~Shojaeifard \emph{et~al.}, ``Fluid antenna systems,'' \emph{IEEE
  Transactions on Wireless Communications}, vol.~20, no.~3, pp. 1950--1962,
  2021.

\bibitem{Wang2025Electromagnetically}
\BIBentryALTinterwordspacing
R.~Wang, P.~Zheng \emph{et~al.}, ``Electromagnetically reconfigurable fluid
  antenna system for wireless communications: Design, modeling, algorithm,
  fabrication, and experiment,'' 2025. [Online]. Available:
  \url{https://arxiv.org/abs/2502.19643}
\BIBentrySTDinterwordspacing

\bibitem{Wong2022Bruce}
K.-K. Wong, K.-F. Tong \emph{et~al.}, ``Bruce {Lee}-inspired fluid antenna
  system: Six research topics and the potentials for {6G},'' \emph{Frontiers in
  Communications and Networks}, vol.~3, 2022.

\bibitem{Zhu2024Historical}
L.~Zhu and K.-K. Wong, ``Historical review of fluid antenna and movable
  antenna,'' \emph{arXiv preprint arXiv:2401.02362}, 2024.

\bibitem{Ying2024Reconfigurable}
K.~Ying, Z.~Gao \emph{et~al.}, ``Reconfigurable massive {MIMO}: Precoding
  design and channel estimation in the electromagnetic domain,'' \emph{IEEE
  Transactions on Communications}, pp. 1--1, 2024, early access.

\bibitem{Zhang2023Analog}
C.~Zhang, S.~Shen \emph{et~al.}, ``Analog beamforming using {ESPAR} for
  single-{RF} precoding systems,'' \emph{IEEE Transactions on Wireless
  Communications}, vol.~22, no.~7, pp. 4387--4400, 2023.

\bibitem{Stoica2008Waveform}
P.~Stoica, J.~Li \emph{et~al.}, ``Waveform synthesis for diversity-based
  transmit beampattern design,'' \emph{IEEE Transactions on Signal Processing},
  vol.~56, no.~6, pp. 2593--2598, 2008.

\bibitem{He2011Wideband}
H.~He, P.~Stoica \emph{et~al.}, ``Wideband {MIMO} systems: Signal design for
  transmit beampattern synthesis,'' \emph{IEEE Transactions on Signal
  Processing}, vol.~59, no.~2, pp. 618--628, 2011.

\bibitem{Zhong2022RMOCG}
K.~Zhong, J.~Hu \emph{et~al.}, ``{RMOCG}: A {Riemannian} manifold
  optimization-based conjugate gradient method for phase-only beamforming
  synthesis,'' \emph{IEEE Antennas and Wireless Propagation Letters}, vol.~21,
  no.~8, pp. 1625--1629, 2022.

\bibitem{Xu20173D}
Q.~Xu, Y.~Huang \emph{et~al.}, ``3-{D} antenna radiation pattern reconstruction
  in a reverberation chamber using spherical wave decomposition,'' \emph{IEEE
  Transactions on Antennas and Propagation}, vol.~65, no.~4, pp. 1728--1739,
  2017.

\bibitem{Costa2010Unified}
M.~Costa, A.~Richter \emph{et~al.}, ``Unified array manifold decomposition
  based on spherical harmonics and 2-{D} {Fourier} basis,'' \emph{IEEE
  Transactions on Signal Processing}, vol.~58, no.~9, pp. 4634--4645, 2010.

\bibitem{Dokmanic2016Sampling}
I.~Dokmanić and Y.~M. Lu, ``Sampling sparse signals on the sphere: Algorithms
  and applications,'' \emph{IEEE Transactions on Signal Processing}, vol.~64,
  no.~1, pp. 189--202, 2016.

\bibitem{Balanis2015Antenna}
C.~A. Balanis, \emph{Antenna theory: analysis and design}.\hskip 1em plus 0.5em
  minus 0.4em\relax John wiley \& sons, 2015.

\bibitem{Zheng2024Mutual}
P.~Zheng, R.~Wang \emph{et~al.}, ``Mutual coupling in {RIS}-aided
  communication: Model training and experimental validation,'' \emph{IEEE
  Transactions on Wireless Communications}, vol.~23, no.~11, pp.
  17\,174--17\,188, 2024.

\bibitem{6758443}
``{IEEE} standard for definitions of terms for antennas,'' \emph{IEEE Std
  145-2013 (Revision of IEEE Std 145-1993)}, pp. 1--50, 2014.

\bibitem{Lu2024Tutorial}
H.~Lu, Y.~Zeng \emph{et~al.}, ``A tutorial on near-field {XL-MIMO}
  communications toward {6G},'' \emph{IEEE Communications Surveys \&
  Tutorials}, vol.~26, no.~4, pp. 2213--2257, 2024.

\bibitem{Grossi2023Beampattern}
E.~Grossi and L.~Venturino, ``Beampattern design for transmit architectures
  based on reconfigurable intelligent surfaces,'' \emph{arXiv preprint
  arXiv:2306.15297}, 2023.

\bibitem{Boumal2023Introduction}
N.~Boumal, \emph{An introduction to optimization on smooth manifolds}.\hskip
  1em plus 0.5em minus 0.4em\relax Cambridge University Press, 2023.

\bibitem{Zheng2024LEO}
P.~Zheng, X.~Liu \emph{et~al.}, ``{LEO}- and {RIS}-empowered user tracking: A
  {Riemannian} manifold approach,'' \emph{IEEE Journal on Selected Areas in
  Communications}, vol.~42, no.~12, pp. 3445--3461, 2024.

\bibitem{Zheng2024Coverage}
P.~Zheng, T.~Ballal \emph{et~al.}, ``Coverage analysis of joint localization
  and communication in {THz} systems with {3D} arrays,'' \emph{IEEE
  Transactions on Wireless Communications}, vol.~23, no.~5, pp. 5232--5247,
  2024.

\bibitem{Selvan2012Descent}
S.~E. Selvan, U.~Amato \emph{et~al.}, ``Descent algorithms on oblique manifold
  for source-adaptive {ICA} contrast,'' \emph{IEEE Transactions on Neural
  Networks and Learning Systems}, vol.~23, no.~12, pp. 1930--1947, 2012.

\bibitem{Alhujaili2019Transmit}
K.~Alhujaili, V.~Monga \emph{et~al.}, ``Transmit {MIMO} radar beampattern
  design via optimization on the complex circle manifold,'' \emph{IEEE
  Transactions on Signal Processing}, vol.~67, no.~13, pp. 3561--3575, 2019.

\bibitem{manopt}
\BIBentryALTinterwordspacing
N.~Boumal, B.~Mishra \emph{et~al.}, ``{M}anopt, a {M}atlab toolbox for
  optimization on manifolds,'' \emph{Journal of Machine Learning Research},
  vol.~15, no.~42, pp. 1455--1459, 2014. [Online]. Available:
  \url{https://www.manopt.org}
\BIBentrySTDinterwordspacing

\end{thebibliography}
\bibliographystyle{IEEEtran}

\end{document}